\documentclass[sigconf]{acmart}

\usepackage{algorithmic}
\usepackage{graphicx}
\usepackage{textcomp}
\usepackage{xcolor, soul}

\usepackage{tcolorbox}
\tcbuselibrary{skins}

\usepackage{booktabs}
\usepackage{longtable}
\usepackage{csquotes}
\usepackage{url}

\AtBeginDocument{%
 }

\setcopyright{acmlicensed}
\copyrightyear{2018}
\acmYear{2018}
\acmDOI{XXXXXXX.XXXXXXX}

\acmConference[Conference acronym 'XX]{Make sure to enter the correct
 conference title from your rights confirmation email}{June 03--05,
 2018}{Woodstock, NY}
\acmISBN{978-1-4503-XXXX-X/18/06}

\begin{document}

\title{Sustaining Maintenance Labor for Healthy Open Source Software Projects through Human Infrastructure: A Maintainer Perspective}

\author{Johan Linåker}

\email{johan.linaker@ri}
\orcid{0000-0001-9851-1404}
\affiliation{%
 \institution{RISE Research Institutes of Sweden}
 \city{Lund}
 \country{Sweden}
}

\author{Georg J.P. Link}
\email{georglink@bitergia.com}
\orcid{0000-0001-6769-7867}
\affiliation{%
 \institution{Bitergia}
 \city{Omaha}
 \state{Nebraska}
 \country{United States}
}

\author{Kevin Lumbard}
\email{kevinlumbard@creighton.edu}
\orcid{0000-0001-9306-3040}
\affiliation{%
 \institution{Creighton University}
 \city{Omaha}
 \state{Nebraska}
 \country{United States}
}

\renewcommand{\shortauthors}{Linåker et al.}

\begin{abstract}
\textbf{Background:}
Open Source Software (OSS) fuels our global digital infrastructure but is commonly maintained by small groups of people whose time and labor represent a depletable resource. For the OSS projects to stay sustainable, i.e., viable and maintained over time without interruption or weakening, maintenance labor requires an underlying infrastructure to be supported and secured. 
\textbf{Aims:}
Using the construct of \textit{human infrastructure}, our study aims to investigate how maintenance labor can be supported and secured to enable the creation and maintenance of sustainable OSS projects, viewed from the maintainers' perspective.
\textbf{Method:}
In our exploration, we interviewed ten maintainers from nine well-adopted OSS projects. We coded the data in two steps using investigator-triangulation.
\textbf{Results:}
We constructed a framework of infrastructure design that provide insight for OSS projects in the design of their human infrastructure. The framework specifically highlight the importance of human factors, e.g., securing a work-life balance and proactively managing social pressure, toxicity, and diversity. We also note both differences and overlaps in how the infrastructure needs to support and secure maintenance labor from maintainers and the wider OSS community, respectively. Funding is specifically highlighted as an important enabler for both types of resources.
\textbf{Conclusions:}
The study contributes to the qualitative understanding of the importance, sensitivity, and risk for depletion of the maintenance labor required to build and maintain healthy OSS projects. Human infrastructure is pivotal in ensuring that maintenance labor is sustainable, and by extension the OSS projects on which we all depend.
\end{abstract}




\maketitle

\section{Introduction}

Open Source Software (OSS) is a pivotal part of modern software supply chains~\cite{blind2021impact}. Today more than 90 percent of proprietary software includes at least one OSS component~\cite{synopsis2020survey, tidelift2018survey}. Much of OSS is community-driven~\cite{capra2008framework}, implying that the OSS is owned and collaboratively developed by a community of individuals that contribute either on their own or on their employer's behalf~\cite{schweik2012internet}. We refer to \textit{the human activity invested by these individuals into the development and maintenance of these OSS projects} as \textit{maintenance labor}. We further make the assumption that OSS projects need a \textit{human infrastructure} to support and secure a sustainable availability of such labor to stay healthy. By human infrastructure, we refer to \textit{the arrangements of organizations and actors within an OSS community that must be brought into alignment, e.g., through governance, processes, and culture, for the OSS project to be viably maintained} (adapted from Lee~\cite{lee2006humaninfrastructure}).

In many OSS projects, large portions of the maintenance labor invested originate from a small core group of community leaders, referred to as maintainers~\cite{wang2020unveiling}. Maintainers typically oversee and help coordinate contributions from the community, performing important maintenance tasks including software development, contributor and user support, and project governance~\cite{eghbal2016roads}. Therefore, maintainers are concerned about building and supporting the human infrastructure needed for the community to engage in open collaboration. Their capacity to maintain OSS projects, including its human infrastructure, varies due to whether the maintenance labor is paid or voluntary~\cite{riehle2014paid}, driven by intrinsic or extrinsic incentives~\cite{lerner2002some}, and the personal situation of the maintainer~\cite{miller2019why}.

Variations and uncertainty about maintainer capacity pose a risk for the ongoing maintenance of an OSS project. For example, increased demand for their time and dedication to their projects may lead to negative stress and burnout, causing them to abandon their OSS projects~\cite{miller2019why}. Sudden life events related to personal interests, family, or working conditions may also impact maintainer capacity~\cite{miller2019why}. Consequently, lower levels of maintainer capacity and inconsistency in maintainer attention to a project may result in development suffering due to lack of community support, decrease in quality of peer-review for potential contributions, and bugs being missed or not prioritized~\cite{eghbal2020working}. From this perspective, the human infrastructure of an OSS project needs to care not only for its community of contributors but also specifically for its maintainer(s) for the project to stay sustainable.

In this study, we aim to qualitatively explore how human infrastructure can help to support the long-term availability of the maintenance labor needed for the project to stay maintained. Specifically, we refer to \textit{an OSS project's ability to stay viable and maintained over time without interruption or weakening} as an \textit{OSS project's health}~\cite{goggins2021open, linaaker2022characterize}. By extension, we make the assumption that for an OSS project to stay or become healthy, there needs to be a sufficient and sustainable amount of maintenance labor available for development and maintenance of the OSS project~\cite{benkler2005common}. If the source of labor is depleted and not renewed, the OSS project may become more or less unmaintained~\cite{linaaker2022sustaining, atkisson2023managing}. Consequences may include introduction of bugs and vulnerabilities, which can have detrimental effects. 

With maintenance labor, we differentiate specifically between labor originating and being added from maintainers (i.e., \textit{Maintainer Labor}) of the OSS project or from the contributors (i.e., \textit{Contributor Labor}) within the community~\cite{schweik2012internet}. As the maintainers are the ones typically building and are responsible for the human infrastructure of an OSS project, we consider their perspective explicitly and define our \textbf{Research Question (RQ)} as:

  \vspace{1mm}
  \textit{From a maintainer's perspective, how can human infrastructure help to secure sustainable availability of maintenance labor from maintainers and contributors for an OSS project to stay healthy?}
  \vspace{1mm}

With sustainable availability, we refer a long-term and replenishing availability of maintenance labor required for an OSS project to stay healthy. To answer the RQ, we deductively designed a questionnaire to interview ten maintainers from nine OSS projects. Through iterative coding we constructed a framework of three high-level themes. Two of these themes relate to the complexities of ensuring the presence of human infrastructure to support Maintainer and Contributor Labor, respectively. The third theme highlights how  labor can be secured or enabled through different funding models. 

\textit{Contributions of our study include} 1) the exploration of OSS project health as dependent on the availability of labor from maintainers (Maintainer Labor) and contributors (Contributor Labor), 2) the characterization of these sources as depletable, 3) a reflection on the fragile relationship between Maintainer Labor and Contributor Labor, and 4) a synthesis of design knowledge, from the maintainer's perspective, that can support the creation and design of human infrastructure to enable sustained availability of maintenance labor. 5) Findings further help maintainers to communicate their needs, and for potential contributors, including organizations, to identify and ask what the needs are so that they can support the maintainer and the OSS projects is the most appropriate way~\cite{microsoftThingsLearned}.

\section{Related Work}

In this section, we provide an overview of the distribution of maintenance effort between maintainers and contributors~\cite{crowston2004effective, mockus2002two, barcomb2018uncovering}. We also describe what factors affect the retention of maintainers, contributors, and newcomers.

\subsection{Distribution of maintenance effort}
The maintenance of OSS projects is commonly performed by a smaller core of contributors, with the assistance and input from episodic contributors and users on the periphery of a project~\cite{crowston2006core}. For example, Avelino et~al.~\cite{avelino2016novel}, found that for 65 percent of their sample, OSS software development was dependant on two or fewer contributors. Similarly, Pinto et~al.~\cite{pinto2016more} observed that episodic contributors make up 49 percent of the total contributor population but their contributions, when measured in code commits, only make up 1.78 percent of the total development effort. 
Despite the seemingly low percentage of development activity, there is a concern, that as the population of episodic contributors in a community grows, the associated workload for the maintainers may also grow, at the expense of planned development~\cite{eghbal2020working, barcomb2020managing}. Wang et~al.~\cite{wang2020unveiling} found that maintainers take on several different tasks, of which coding is just a limited portion, and the more mature a project is, the more effort is spent by maintainers on supportive and communicative activities. They concluded that \textit{``efforts in non-technical activities are negatively correlated with the project’s outcomes in terms of productivity and quality in general, except for a positive correlation with the bug fix rate (a quality indicator)''}. Pinto et~al.~\cite{pinto2016more} noted that episodic contributions may bring value, including non-trivial contributions, but come at the cost of time that is required to review and assist them through the contribution process. Barcomb et~al.~\cite{barcomb2020managing} also found that episodic contributors come with an associated cost and concerns for quality. Linåker \& Runeson~\cite{linaaker2022sustaining} note that the demand for labor, required to maintain and manage contributions, increases as a projects' adoption and popularity grow.

\subsection{Attracting and retaining episodic contributors}
Attracting and retaining episodic contributors, and thereby addressing discrepancies in project knowledge and difficulties in knowledge exchange between maintainers and episodic contributors is difficult.~\cite{barcomb2020managing}. Episodic contributors may not have the same motivation, nor time to dedicate to the project as its maintainers~\cite{pinto2016more, iaffaldano2019developers}. Episodic contributors often also face longer time in the review process and lower acceptance rates which may negatively affect their motivation to contribute~\cite{lee2017one}. Lowering barriers to entry~\cite{steinmacher2019overcoming} and creating on-boarding processes for newcomers~\cite{steinmacher2014attracting} have been identified as an important means of reducing maintainer review efforts, while increasing acceptance rates of contributions and increasing retention of diverse contributors~\cite{fondhjem2021onboarding, yue2022off}. Contribution barriers can be contextualized at the foundation, project, and individual level regarding contributing and engaging with a community~\cite{guizani2021long}. Contribution barriers for newcomers and episodic contributors can be both social, technical, and process-related~\cite{steinmacher2019overcoming, constantino2023perceptions}. 

\subsection{Retaining maintainers and core contributors}
While attracting and retaining episodic contributors is important to sustain and grow a project, much of the key work in projects are done by maintainers and core contributors~\cite{avelino2016novel,pinto2016more}. Thus, retaining maintainers and core contributors of an OSS project is critically important for a project. Constantinou et~al.~\cite{constantino2023perceptions} highlighted the risk of overwork that may lead to maintainers abandoning a project and also how financial incentives may enable maintainers to dedicate more time to a project, explored further in several studies( e.g., ~\cite{overney2020how, zhou2022studying, riehle2012single}). Iffaldano et~al.~\cite{iaffaldano2019developers} found that reasons for abandoning a project may be personal (e.g., relating to the contributors' professional situation, life-changing events, changes to the financial situation) or project-related (e.g., related to a role change, the current social situation in the project, technical or organizational changes in the project, or a feeling that there is no need to contribute anymore). In a similar study, Miller et~al.~\cite{miller2019why} identified three common reasons for leaving an OSS project: occupational (e.g., the maintainer getting a new job or role that does not support OSS-related work, dropping out of school, or not having time left to work on the project), social (e.g., lack of interest, no personal time, or lack of peer support), and technical (issues with collaboration tools or the industry in general). Zhang et~al.~\cite{zhang2022turnover} confirmed these findings in a study investigating why companies leave OSS projects.

\section{Research Design}

We conducted an exploratory qualitative interview survey with maintainers from a sample of OSS projects with overlapping and distinctive characteristics. Below we describe how we 1) deductively constructed a questionnaire, 2) conducted a series of interviews, and 3) coded the transcripts in a two-step process. This synthesis resulted in a framework of infrastructure design aspects to consider when building human infrastructure that can help to enable a healthy OSS project, which we introduce in Section~\ref{sec:Findings and Analysis - The OSS Resources Framework}.

\begin{table*}[!tbh]
\caption{Overview of maintainers and their OSS projects. *Backed by a software vendor. **Backed by a foundation.}
\label{tab:maintainers}
\begin{tabular}{ p{0.5cm} p{3cm} p{3.8cm} p{1.5cm} p{1.1cm} p{1.5cm} p{1.5cm} p{2cm} }
\toprule
\textbf{ID} & \textbf{Type} & \textbf{Function} & \textbf{Language} & \textbf{Started} & \textbf{Maintainers} & \textbf{Committers} & \textbf{License}\\ \midrule
M1 & Middleware      & FTP/HTTP client        & C       & 1996 & 1 & 39  & BSD-3/4-clause\\
M2 & Middleware      & FTP/HTTP client        & C       & 1996 & 2 & 3   & GPL 3 or later\\
M3 & Server application  & Web server          & Go      & 2015 & 1 & 30  & Apache 2.0\\
M4 & Middleware /Library  & Image manipulation      & C       & 1990 & 2 & 25  & Project-specific\\
M5 & Middleware      & FTP/HTTP client        & C       & 1996 & 2 & 3   & GPL 3 or later\\
M6 & Webb application   & Computer-assisted translation & Python    & 2012 & 2 & 60  & GPL 3 or later\\
M7 & Cloud infrastructure & Orchestration engine     & Python/Ruby & 2012 & * & 33  & GPL 3\\
M8 & Javascript library  & Content management      & Javascript  & 2015 & 1 & 10  & GPL 3\\
M9 & Desktop application  & Office productivity suite   & C++      & 2011 & ** & 58  & MPL 2.0\\
M10 & Desktop application  & VPN client          & Python    & 1998 & 1 & 9   & LGPL 2.1\\ \bottomrule
\end{tabular}
\end{table*}

\subsection{Questionnaire Design}
To limit researcher bias when constructing the interview questionnaire, we followed the assurance case method. It offers a formal and auditable means of deductively developing an interview questionnaire from an initial problem understanding~\cite{gandhi2009assurance}. It motivates each question and helps improve the construct validity by ensuring that each question is relevant and addresses the research questions. 

\begin{figure}[h!]
  \centering
  \includegraphics[width=0.85\columnwidth]{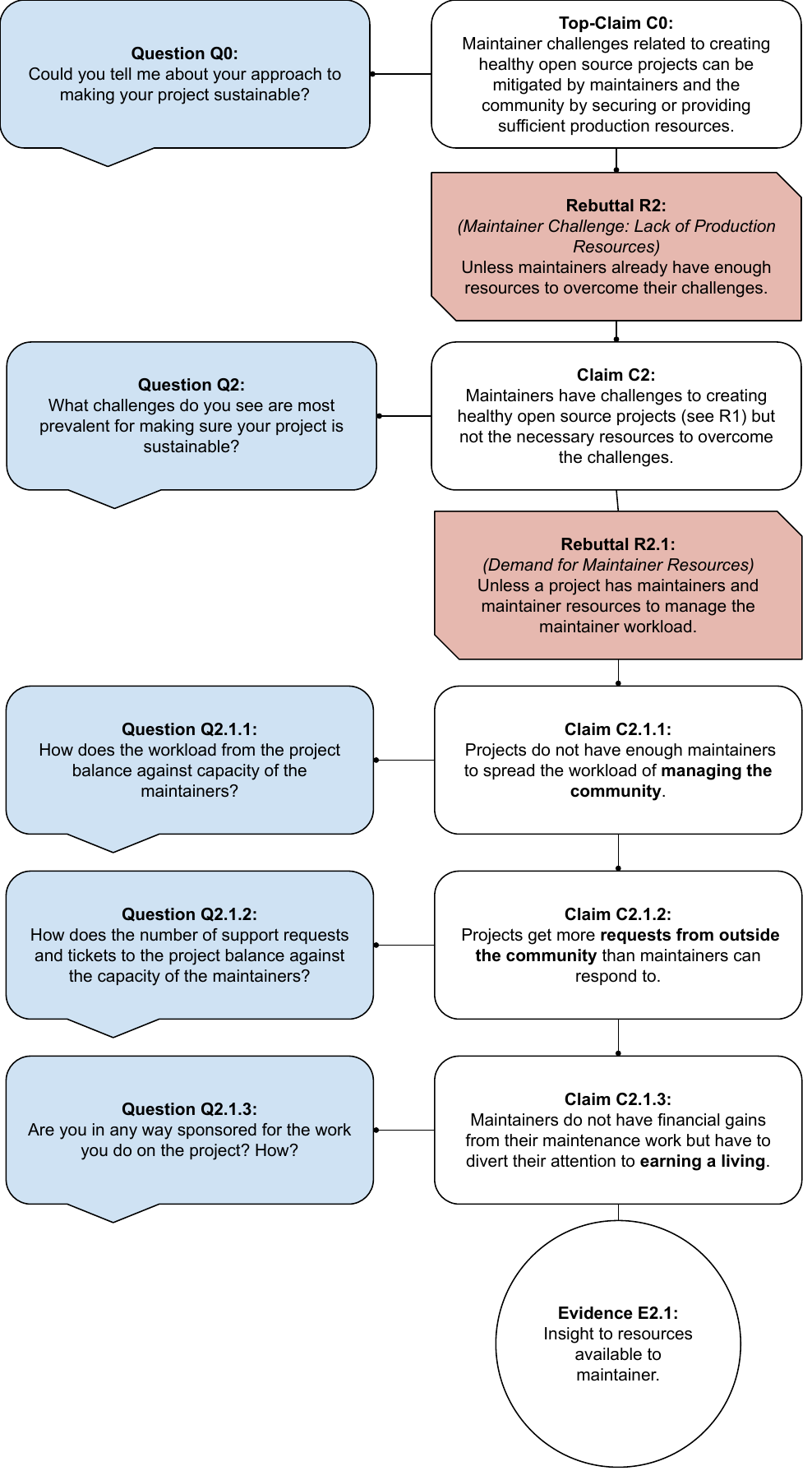}
  \caption{Sample branch of the visual assurance case starting with a top-claim, followed by rebuttals and sub-claims until ending in an evidence statement, highlighting what information needs to be collected to affirm overarching claims.}
  \label{fig:visual-assurance-case}
\end{figure}

Our process was as follows. From our initial assumption and informed by our understanding of the problem context, we defined our top-level claim that challenges related to creating a healthy OSS project can be mitigated by securing or providing sufficient levels of maintenance labor, either from maintainers or contributors (see supplementary material for a copy of the visual assurance case~\cite{SupplementaryMaterial}). Rebuttals were then defined to cast doubt on the validity of the top-level claim, e.g., \textit{...unless maintainers already have enough resources to overcome their challenges}. The intentional formulation of doubts as rebuttals addresses the concern for confirmation bias and blind spots in constructing the interview questions. For each rebuttal, a series of sub-claims are defined, again informed by literature and our initial problem understanding, that strengthen the top-level claim. 
Each branch with a rebuttal and derived set of sub-claims was summarized with a statement of what evidence would need to be collected to affirm the sub-claims and reject the rebuttal of the branch. Questions were then drafted based on each sub-claim, focusing on the information needed to affirm or reject the sub-claim and, by extension, its overarching rebuttal. To avoid the risk of leading questions, these were formulated as open-ended and inspired by example formulations by Spradley~\cite{spradley2016ethnographic}. For a sample branch of the assurance case used in this study and the related questions used in our questionnaire, see Fig.~\ref{fig:visual-assurance-case}.

We used the assurance case also to validate the questionnaire with two experts within OSS, one from industry and one from academia. The assurance case offered a deeper conversation within the research team during the development and with the experts during validation, much deeper than would have been possible by only reviewing the list of questions. The assurance case was not used after the development of the questionnaire, which is available in the supplementary material~\cite{SupplementaryMaterial}. A pilot interview was also performed with an OSS developer, which rendered slight adjustments to the phrasing and to the way the concept of OSS project health was introduced during the interviews. 

\subsection{Data Collection: Interviews}
In our investigation, we were specifically interested in the population of OSS projects that are maintained by two or fewer maintainers. We assume that these are more susceptible to challenges in regard to securing a sufficient and sustainable amount of maintenance labor necessary for a healthy OSS project. We do not intend to generalize our findings to specific types of OSS projects, rather we are focus on exploring how human infrastructure can help to secure a sustainable availability of the Maintainer and Contributor Labor needed for an OSS project to stay healthy. We choose to explore the maintainer perspective explicitly as we consider them vulnerable to the potential impact of limited maintenance labor, but also as they may be considered most knowledgeable about their own OSS projects, and the challenges they face.

We interviewed ten maintainers representing nine different OSS projects, as presented in Table~\ref{tab:maintainers}. Of these, seven maintainers from six OSS projects fit into this profile. To provide contrast, we sampled two maintainers from a large corporately-sponsored OSS project and a large community-driven OSS project. Maintainers M1, M4, and M9 were selected through targeted sampling within our existing networks. Remaining maintainers were sampled based on recommendations from the first set of maintainers.

Interviews were conducted online and recorded. The first and second authors both attended the majority of the interviews, where they took turns asking questions and taking notes. Interviews lasted between 30-60 minutes and were semi-structured following the questionnaire. The authors bring over two decades of combined experience from actively participating in open-source communities, conducting OSS research, and working in the software development industry. This deep well of knowledge allowed us to understand the nuances of conversations and delve deeper with targeted questions. For instance, we could inquire further when a maintainer offered specific examples or relevant personal experiences.

Audio recordings were automatically transcribed using software tool support. The raw transcripts were edited to provide consistent structure, corrected for grammar and misspellings, and anonymized. The processed transcripts were then sent to maintainers for validation and confirmation to ensure correctness and offer the opportunity to remove sensitive information.

After five interviews, the first and second authors performed a summary of findings based on personal note-taking, transcripts, and general reflections. The purpose was to gain an overview, improve understanding, and gauge saturation in the data. After four additional interviews, the first and second authors revisited the summary of findings and could confirm that statements were being repeated and that new insights gained per interview were limited.

\subsection{Data Analysis}
Data was analyzed in a two-step coding process, as explained below.

\subsubsection{Analysis Step One - Structured, Simultaneous, and Open Coding:}
We began with a structured coding approach in which we categorized and structured our data according to a set of codes~\cite{saldana2021coding}. Codes were divided into three main categories challenges (in terms of securing the availability of the necessary maintenance labor), impact (of the challenges), and actions (that may address the challenges and reduce their impact). Specifically, an a-priori codebook (see~\cite{SupplementaryMaterial}) was developed from the questionnaire, grouping answers across interviews into the aspects needed for our research question.

The transcripts were formatted in a paragraph structure and coded individually by the first and second authors. Following the technique of simultaneous coding, also known as double coding, we assigned relevant codes to the paragraphs, often more than one code. This enabled us to maintain the connection between codes and investigate patterns of co-occurrences between different codes.

New codes were introduced through open coding when new concepts and ideas were identified. Existing codes were merged when appropriate. After each interview, the first and second authors convened and discussed their coding of each paragraph, taking turns in starting the discussion. The third author participated in the discussion of the first five interviews as a judge for settling disagreements or by adding a third perspective when the first and second authors requested. For the remaining five interviews, the first and second authors mutually settled disagreements.

\subsubsection{Analysis Step Two - Axial and Selective Coding:}
In this second analysis step, the first author conducted axial coding by looking at all the paragraphs within each code and synthesizing key ideas into short descriptive sentences about that code. The technique was chosen to inductively build an understanding of the data~\cite{saldana2021coding} in the challenges and actions code categories. We decided not to code the impact codes separately because they always co-occurred with the challenges codes. The descriptive sentences from separate interviews and paragraphs were then grouped to arrive at recurring themes through selective coding~\cite{saldana2021coding} highlighting the infrastructure design aspects as reported in Section~\ref{sec:Findings and Analysis - The OSS Resources Framework}. The third author performed a second iteration on the selective coding to fine-tune and validate the final coding scheme. An audit trail was maintained throughout the process, enabling the second author to review and validate the trace from paragraphs to the recurring themes. Finally, the constructed themes, conceptual narratives, and disagreements were discussed and addressed together by all authors. 

\section{Findings and Analysis}
\label{sec:Findings and Analysis - The OSS Resources Framework}
\begin{figure*}
  \centering
  \includegraphics[width=1\linewidth]{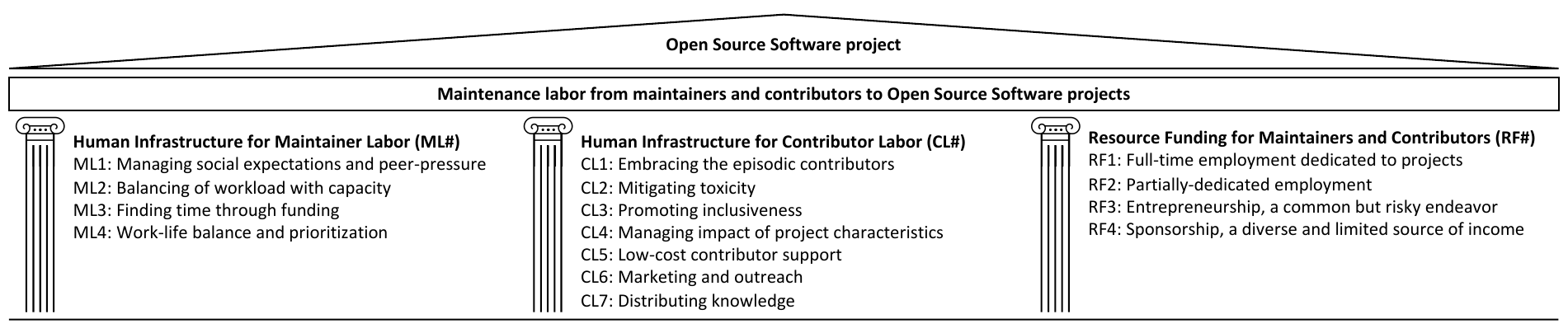}
  \caption{Overview of the human infrastructure design aspects to consider in process of enabling and securing the necessary amount of labor from maintainers and contributors for the OSS project to stay or become healthy.}
  \label{fig:human-infra-design-aspects}
\end{figure*}

The section presents a framework of infrastructure design aspects (see Fig.~\ref{fig:human-infra-design-aspects} to consider when building human infrastructure to help enable and support a healthy OSS project. The aspects are divided into three categories. \textit{Maintainer} and \textit{Contributor Labor} include aspects highlighting the opportunities and complexities of ensuring the presence of the necessary maintenance labor from maintainers and contributors of an OSS project, respectively. \textit{Resource Funding} includes aspects of how these resources may be funded.

\subsection{Maintainer Labor}
Maintainers have a responsibility to the OSS community to ensure the continued development of quality software. However, without adequate time, maintainers can find it difficult to meet community expectations. Below, we present infrastructure design aspects (ML\#) of human infrastructure that maintainers may consider to ensure a sustainable availability, specifically of Maintainer Labor for their OSS project. Each aspect is concluded with a brief summary.

\subsubsection{ML1: Managing social expectations and peer-pressure}
Maintainers can experience expectations and a form of social pressure. In part, this relates to a self-reflected implicit feeling of responsibility towards the user base that depends on the project, a feeling that may grow along with the community.

\vspace{1mm}
  \textit{``I suppose, part of [the pressure] is driven by social expectation. [...] You feel pressured to improve your software because people complain instead of contributing. And that is human nature, I guess.''} - M3
\vspace{1mm}

A second part of this social pressure is related to communication processes where there is an explicit expectation from the user base of timely responses from maintainers to their inquiries and solutions to their requests. OSS communities expect quick solutions when they are experiencing critical issues. Expectations of maintainers, both perceived and experienced, add stress and pressure on maintainer time as a resource. Maintainers underline that their time is a finite resource that needs to be protected to be sufficiently available for the expectations of a growing community. 

\begin{tcolorbox}
Expressed or perceived expectations from companies, users, or developers on maintainers' support may risk impacting negative stress.
\end{tcolorbox}

\subsubsection{ML2: Balancing of workload with capacity}
Maintainer time as a resource associated with Maintainer Labor is impacted by workload, which is the amount of work and difficulty of the work in an OSS project. Maintainer interactions with an OSS projects' community take up a large amount of time and are often prioritized over software development. The community interaction grows with the number of issues and users of the project, as does the technical maintenance required due to a growing code base due to contributions continuing to come in.

\vspace{1mm}
  \textit{``The maintenance burden falls on us, especially with [episodic] feature contributions. Sure, it is great [...], but that just means there is more code that I have to maintain.''} - M5
\vspace{1mm}

Maintainers may, therefore, experience an increased workload as the community grows. If the workload exceeds their capacity to work on the project, consequences may include a negative impact on maintainer time as a resource, which can lead to a decline in the project's development velocity and quality. Further, maintainers may have less capacity to dedicate to general feature development and improvement of project artifacts such as documentation.

\begin{tcolorbox}
Amount of perceived work for maintainers generated from the OSS community (e.g., support requests, bug fixes, community management) need to balance against maintainers' capacity.
\end{tcolorbox}

\subsubsection{ML3: Finding time through funding}
Maintainers who are paid to work on a project can allocate more time to the project and are empowered to reconsider work-life balance when the maintenance work helps support their lifestyle. Increasingly, maintainers believe there needs to be monetary support or incentives to prioritize work on projects. Interviewees note that motivation is not necessarily altruistic and that they prefer being paid. 

\vspace{1mm}
  \textit{``I think OSS is not itself a virtue.''} - M3
 \vspace{1mm}

However, even when a maintainer is working in a paid capacity, work has a habit of overflowing into their personal time due to community expectations. 

 \vspace{1mm}
  \textit{``It could become a pressure for people trying to balance [maintenance] with their normal life and workload. I get that input from a lot of maintainers... You want to earn money, and you want to work on your OSS project, but you still do not want to do too many hours each day.''} - M1
\vspace{1mm}

For key OSS infrastructure projects, the paid capacity of maintainer time has become the expectation. This expectation is the result of increased organizational engagement with OSS projects, where there are expectations of sustainability for the software organizations' needs. 
\begin{tcolorbox} Funding may help maintainers dedicate more professional time towards a project, while prioritization may still be an issue.
\end{tcolorbox}

\subsubsection{ML4: Work-life balance and prioritization}
While an increasing number of maintainers are receiving some form of compensation for their work, many maintainers are still not paid to work on their projects. Thus, work is often done in their spare time. They feel a passion for their projects and explore different approaches to be able to work on them in their spare time without financial support. Managing maintainer time that creates a positive work-life balance is difficult. One maintainer described a reluctance to allocate spare time to project work:

\vspace{1mm}
  \textit{``My priorities in my free time are constantly changing, and I do not always want to allocate my time on this one project, particularly when I do not use it myself.''} - M10
\vspace{1mm}

Having clear and explicit personal and work task prioritization may help the maintainers stay focused. This is, however, not perceived as an easy task for maintainers as certain issues and the timing of those issues can not be controlled. Security-related issues are one example that maintainers prioritize when they emerge. 

\begin{tcolorbox}
Creating a work-life balance requires personal prioritization by the maintainer over their perceived or experienced need to maintain their projects.
\end{tcolorbox}

\subsection{Contributor Labor}

In addition to maintainers' time, OSS projects are the result of labor from many different types of contributors. Contributors are the lifeblood of any OSS project. Contributor labor is a resource that needs to be supported and managed in sustainable communities. As such, design aspects of human infrastructure focusing on \textit{Contributor Labor} consider how to enable the attraction and retention of contributors providing high-quality contributions to the OSS projects. Each design aspect (CL\#) is presented below and concluded with a brief summary.

\subsubsection{CL1: Embracing the episodic contributors}
Interviewees report that contributions are commonly made by episodic contributors, who, when they contribute, address a specific issue representing a personal or organizational need. When this is addressed and implemented, they move on. Retaining these episodic contributors or turning them into core contributors is perceived as difficult. 

\vspace{1mm}
  \textit{``We accept that people come and go. I used to get a bit worried but now I just accept that this is the way of the world.''} - M7
\vspace{1mm}

By accepting the continued presence of episodic contributors, maintainers may consider focusing on how to best enable their contributions to be of as high quality as possible rather than how the contributors can be retained.

\begin{tcolorbox}Episodic contributors should be encouraged and empowered to make their contributions as high quality as possible.
\end{tcolorbox}

\subsubsection{CL2: Mitigating toxicity}
Several interviewees reported experiences of toxicity in discussions and communications in their OSS projects, e.g., through aggressive behavior by pull request authors. 

\vspace{1mm}
  \textit{``It used to be pretty much that it was a full sort of technocrat sort of thing. Everything is about making the other person look stupid. [...] You would have a code change [for a plugin]. You would get shouted at for doing something that the author does not like''} - M8
\vspace{1mm}

The toxicity also comes from users of the OSS who complain about bugs and issues in a way that puts negative stress and pressure on maintainers and other contributors. In cases where the project is closely linked to a single or primary maintainer, he or she may become the focal recipient for harsh and unfiltered feedback, which can have a personal impact on the maintainer. Enforcement of culture, e.g., through a code of conduct and proactive responses, was highlighted practices.

\begin{tcolorbox}Toxicity is commonly experienced and requires a proactive approach and building of a positive culture enforced, e.g., through a code of conduct.
\end{tcolorbox}

\subsubsection{CL3: Promoting inclusiveness}
Growing and fostering a positive and inclusive culture was highlighted by maintainers as pivotal to growing the community and encouraging contributions. Specifically, maintainers highlighted the need to listen and discuss openly where the community wants to go with the project, and thereby make the community feel included in the decision process.

\vspace{1mm}
  \textit{``My efforts around [sustainability] is to make sure to be open and welcoming by encouraging others to take part, responsibility, and to share the burden... We are open to giving people responsibility within the project. It doesn't have to be me doing things.''} - M1
\vspace{1mm}

The maintainers exemplified mentorship and on-boarding processes as means of actively welcoming new contributors, but also the need to encourage, be responsive to, and show appreciation for new contributors.

\begin{tcolorbox} A Sense of inclusiveness is considered pivotal for new contributors to stay around. Highlighted means include encouragement and appreciation for contributions, openness in decision-making, and responsiveness in answers.
\end{tcolorbox}

\subsubsection{CL4: Managing impact of project characteristics}
Some maintainers noted technical aspects are likely factors for the lack of contributors. People may be less aware of projects located in lower or middle layers of technology stacks or be less interested in projects based on older technology. The complexity of the code may also be a related rationale.

\vspace{1mm}  
  \textit{``We are also a project that is old and written in C and sometimes complicated stuff. So, it also limits the number of developers we actually attract and who would even consider contributing.''} - M1
\vspace{1mm}

Another aspect of representing a stable and established project is that users are accustomed to how it works and are, hence, less willing to contribute or pay for support. It may also be that they assume that the project has all the help it needs or does not need any as it appears stable and has been around for a long time. In these cases most feature requests may concern edge cases but maintainers stress that OSS projects still need to adapt to surrounding technologies and protocols which require both time and resources.

\begin{tcolorbox}Awareness and actions are needed to address the potential negative impact of a project's characteristics (e.g., programming language, position in the stack, and general stability) on the attractiveness of new contributors.
\end{tcolorbox}

\subsubsection{CL5: Low-cost contributor support}
Lack of knowledge and experience in technical and legal matters among users and (potential) contributors was expressed as a complex challenge. Unfamiliarity with, e.g., the infrastructure or development process of an OSS project may impose a barrier to contribution while also leading to low-quality contributions from those who pass the initial barriers. According to interviewees, this is evident in questions that are of too simple a nature, bug reports that lack important details, and pull requests that are too big and overly complex. Maintainers are facing a balancing act between lowering contribution barriers to promote contributions and maintaining quality when managing Contributor Labor. 

\vspace{1mm}
  \textit{``In lowering the barrier for contributions, we invited a lot of low-quality contributions and didn't seem to expect much from our users. If we're going to put our best effort into maintaining, we want our users to put in their best effort, too. That's the only way that even remotely works.''} - M3
\vspace{1mm}

The lack of certain types of contributions and the sometimes poor level of quality add to the workload of the maintainer impacting them both on a personal level, as well as on the project' development velocity and overall quality. Hence, maintainers need to reduce barriers to entry and provide support that raises contribution quality without consuming the time that is available. Project documentation, onboarding programs for new contributors, mentoring, and training contributors are some proposed means.

\begin{tcolorbox}Contribution barriers need to be removed while support measures help increase the quality of contributions without consuming too much of the maintainers' available time.
\end{tcolorbox}

\subsubsection{CL6: Marketing and outreach}
Marketing was highlighted as a key practice for attracting project funding, new contributors, and generally growing the community. However, multiple maintainers expressed discomfort with the activity and lack of time or knowledge as reasons for not prioritizing marketing. Many maintainers, therefore, explicitly expressed a wish for new contributors and support focused on helping them market their projects. One maintainer discussed the need for a dedicated marketing and community outreach position in their project. 

\vspace{1mm}
  \textit{``What we definitely need is someone who says: `Okay, I will help you advertise all this stuff`... My motivation is more or less purely technical, and of course, contact with the user base is very motivational. But doing advertising is not my world.''} - M2
\vspace{1mm}

Although maintainers have mixed feelings, they indicated some confidence in promoting the project indirectly in more technically oriented contexts, including talking and networking at events such as conferences, meetups, and hack-a-thons about the project, the roadmap, and how others can help.

\begin{tcolorbox}Maintainers need help with marketing and outreach, as they may feel uncomfortable and prefer to focus on technical aspects of their projects.
\end{tcolorbox}

\subsubsection{CL7: Distributing knowledge}
Several maintainers amplified the idea that they should try to share as much knowledge as possible and avoid any secret or tacit knowledge, thereby decreasing any dependencies on them should they go away. One maintainer notes:

\vspace{1mm}
  \textit{``Since I think we have most of everything documented, everything open, everything is sort of free and everywhere, my hope or intention is that if I drop out tomorrow---oh sure, it will be some sort of bump in the road, but the project will survive. Everything is out there. I do not have any secrets. I do not have any magic keys to the kingdom.''} - M1
\vspace{1mm}

Technical and user documentation was, accordingly, highlighted by several as key to capture and communicate knowledge about the project, especially on how to contribute and engage in the project. Documentation should preferably also cover the long-term planning of the project so people know what to expect and what can be contributed, and where they can help out. 

\begin{tcolorbox}Sharing knowledge in communication and documentation decreases dependency on the single maintainer while improving onboarding and inclusiveness.
\end{tcolorbox}

\subsection{Resource Funding}
Funding can enable the provisioning of maintenance labor required to maintain a sustainable OSS project. OSS development has seen increasing amounts of corporate engagement, and maintainers are coming up with new and novel ways of supporting their projects. Paid capacity for maintainers and contributors has become the new normal for large and critical projects. However, attaining and managing project funding can be difficult for maintainers of projects that don't have support from large companies. Smaller and medium-sized projects have begun to introduce and utilize new and novel approaches to attract and manage monetary resources in kind. Aspects related to funding (RF\#) are listed below.

\subsubsection{RF1: Full-time employment dedicated to projects} 
Being employed to maintain or contribute to an OSS project is no longer considered a rare occurrence, especially for maintainers and contributors of larger projects. For smaller projects, as mainly represented by the interviewees, it is reported as less common but is highly desirable. Employment is seen to relieve stress compared to self-employment, as the maintainer would get benefits, such as medical insurance, in many countries.

\vspace{1mm}
   \textit{``If you are an employee instead of self-employed, then you can get [social] benefits. And that takes the stress levels way down. [...] You don't have to worry about your day-to-day like, Oh man, like should I go make dinner or should I work on this project?''} - M3
\vspace{1mm}

Increasingly, companies are viewed as willing to pay for full-time employment of contributors to projects that are critically important to their organizations. Another rationale is that the company provides a commercial support offering for an OSS project to its customers. A perceived risk with this model was that the maintainer might be put to work on things that are beneficial for the company and their customers rather than the community as a whole.

\begin{tcolorbox}Full-time employment dedicated to working on projects is seen as a dream by many, especially if social benefits (e.g., medical insurance) are included.
\end{tcolorbox}

\subsubsection{RF2: Partially-dedicated employment} 
Some companies allow maintainers or contributors to spend part-time on an OSS project used by the employer. For important projects, the rationale may include cost savings through shared maintenance or support of the project and build influence. Interviewees, however, often referred to possibilities of being able to work on projects (as their own) that were not necessarily used by their employer. 

\vspace{1mm}
  \textit{``You have a few hours per week to work on your favorite projects. This is a kind of further education for people. I brought this kind of knowledge into my company, like GIT, which we didn't use.''} - M2
\vspace{1mm}

Many of the interviewees describe both in terms of training for themselves, knowledge transfer to the employer, and a personal benefit for the employee.

\begin{tcolorbox}Being able to spend part-time employment to maintain a project is viewed favorably as a means of personal training and benefit while bringing value to the employer through, e.g., knowledge transfer.
\end{tcolorbox}

\subsubsection{RF3: Entrepreneurship, a common but risky endeavor} 
Building a business was reported as a popular but difficult alternative to employment. A common model regards offering commercial support and custom development to the OSS project, either through a larger business consultancy or as a freelancer. The latter was considered riskier and more of an opportunity for smaller and occasional contracts than supporting a livelihood. Wrapping proprietary functionality around the OSS project (cf. open core) and providing support and hosting services were also variants used. 

Another model keeps everything in the open, and paying customers directly influence the prioritization of the project's roadmap. This was, however, a model viewed from different angles:

\vspace{1mm}
  \textit{``A customer is paying for something that would be developed in the future, but it will be developed sooner thanks to them, and it will help [my project] to be sustainable.''} - M6
\vspace{1mm}

Another maintainer noted that it would be difficult to prioritize community issues or answer questions when demand is high from paying customers and available resources are low. If the community is not given enough attention, it may, e.g., risk impacting contributor growth and retention. On the other hand, when a maintainer can separate the business demands from impacting the steering of the project, it can enable others to also provide similar services, thereby benefiting the project and the maintainer in the long run. 

\begin{tcolorbox}Building a business comes with many variants and is considered risky to different extents. Prioritizing the needs of paying customers and the community requires a delicate balance.
\end{tcolorbox}

\subsubsection{RF4: Sponsorship, a diverse and limited source of income}
Sponsorship as a source of income can be differentiated between private sponsorship and corporate sponsorship. Private sponsorship, that is, money donated by individuals, was generally considered to have a limited effect on project sustainability as the sums are often very small. A large following would be needed for private sponsorship to reach larger sums and have a stronger impact. However, most maintainers consider the sponsorship, although small, as a form of recognition and motivational booster. Maintainers consider it symbolic ``coffee money'' and sometimes it allows them to buy hardware for the project or pay for travel.

\vspace{1mm}
  \textit{``For me, it is pocket money because it is around a hundred dollars a month. It is not something you can live off of. It is just something extra... I don't have a clear goal because the amount is not high. If it was enough, I'd work less.''} - M4
\vspace{1mm}

Corporate sponsorship was considered the preferred model as it can have a substantial impact through potentially larger donations. Maintainers highlighted that corporate sponsorship could come in non-financial forms, such as a free SaaS subscription, infrastructure (for hosting, testing, building), or marketing of the OSS project. To enable and receive corporate sponsorship, maintainers may need a legal organization that can manage any funds. 

\begin{tcolorbox}Personal sponsorship is considered a symbol of gratitude, while corporate sponsorship is more significant. None is, however, considered a sustainable source of income.
\end{tcolorbox}

\section{Discussion}
Below, we discuss our findings and provide propositions for future work to explore and validate further.

\subsection{People as a depletable resource}
This study provides an empirical and exploratory perspective on the concept of OSS project health and sustainability from a human infrastructure perspective. Specifically, how does human infrastructure support the sustainable availability of maintenance labor needed to ensure healthy OSS projects. The infrastructure design aspects identified highlight that human infrastructure in OSS is not just about platforms, processes, and governance structures, but is equally about the human factors in play among the people performing the development and maintenance of the OSS projects. 

People are at the front and center of OSS projects' underpinning open and ideally collaborative development. Their time and labor are scarce resources that can be depleted if not cared for or replenished, as with common pool resources. Atkisson and Bushouse, in similar veins, describe the time and labor from the wider community as volunteer energy and how this also needs to be sustained to prevent depletion~\cite{atkisson2023managing}. Here, we differentiate between ``energy'' or labor coming from the maintainer(s) (Maintainer Labor) or the contributors (Contributor Labor). We find that these two sources sometimes require different and sometimes overlapping support from the human infrastructure to be sustained. Further, we address the changing nature of OSS as it skews away from a primarily volunteer activity to an increasingly paid activity.

\subsection{Balancing influx of Maintainer and Contributor Labor}
These two resource types represent complementary sources of input to the human infrastructure required to sustain the development of an OSS project. What we also note is a dependency between the two, where too much or less of one or the other can have a negative impact on the OSS project's health. A non-existent or limited pool of contributors may lead to a cumbersome burden and dependency on the maintainers. On the other hand, observations from our interviews tell us (in line with extant research~\cite{linaaker2022sustaining, eghbal2020working, barcomb2020managing, wang2020unveiling, pinto2016more}) that as the OSS project and its community grows, so does the maintenance burden on the code base, and the overhead required to manage the increasing amount of community interaction that is demanded. This demand is experienced both implicitly among the maintainers as a feeling of responsibility and explicitly from the user base who expect timely responses, a frequently identified barrier-to-entry for newcomers~\cite{guizani2021long, steinmacher2019overcoming}. The community interaction typically relates to feature requests/bug reports, suggestions for corrections, or voicing an opinion. Limited and low-quality contributions, along with limited technical knowledge among the (potential) users and contributors, were also highlighted as challenges.

Attracting and retaining Contributor Labor should, hence, (ideally) be enabled in such a way that they are welcomed and inclined to stay while also enabled and onboarded in such a way that their contributions come with a higher level of quality and lower level of attention required from the maintainers. In other words, Contributor Labor must not come at the expense of too much Maintainer Labor, an equation that is difficult to solve as there will always be a need for maintainers to care for the community and facilitate an open collaboration. Per the interviewees, this was, however, not seen as a need but rather as a will and desire, as everyone is keen on spreading the adoption of their OSS projects.

\begin{tcolorbox}\textbf{P1}: As the use and popularity of an OSS project grow, so does the demand for Maintainer Labor to manage non-coding tasks, such as answering questions, triaging bug requests, and reviewing pull requests.
\end{tcolorbox}
\begin{tcolorbox}\textbf{P2}: As the use and popularity of an OSS project grow, so does the perceived and expressed social pressure on the maintainers to dedicate more time to the maintenance of the OSS project.
\end{tcolorbox}

\subsection{Attraction and onboarding of Contributor Labor}
Several ways of attaining this balance between attraction, retention, and contribution quality were suggested. One such way concerned the aspect of marketing and outreach, something that the maintainers interviewed did not feel comfortable with or were willing to prioritize in contrast to technical work, which their interest generally leaned towards. Non-coding contributions in the form of marketing, support, and community management may, hence, provide a means of relieving the stress of maintainers while enabling them to dedicate more time to development-related tasks (cf.~\cite{fang2022damn}).

The removal of barriers for newcomers was also highlighted both by interviewees and literature~\cite{steinmacher2019overcoming, guizani2021long, constantino2023perceptions}. This includes providing clear directions about the current focus and direction of the project, which may help to reduce questions and unnecessary contributions. Pointing out issues that a project needs help with and arranging dedicated community events can be other means of directing potential contributors to prioritized areas of a project. Addressing common technical barriers~\cite{steinmacher2019overcoming, guizani2021long, barcomb2020managing}, e.g., through documentation and standardized infrastructure, may reduce the need for technical support. Growing a positive and welcoming culture and atmosphere within the projects may help increase willingness to continue engaging in the project.

\begin{tcolorbox}\textbf{P3}: Non-coding contributions in the form of marketing, support, and community management relieve the stress of maintainers and enable them to dedicate more time to development-related tasks.
\end{tcolorbox}
\begin{tcolorbox}\textbf{P4}: Demand for Maintainer Labor is reduced if contribution quality is increased, e.g., by reducing barriers for newcomers and onboarding that convey the necessary technical and process-related knowledge.
\end{tcolorbox}

\subsection{Resource Funding enables healthy developers and projects}
While the attraction, retention, and onboarding of contributors provide one lever, \textit{Resource Funding} was unanimously flagged as an enabler for both the inflow of Maintainer and Contributor Labor. Dedicated funding may allow maintainers and contributors to dedicate time to their projects while still allowing for a work-life balance, reported as a key challenge, aligning with others' observations~\cite{constantino2023perceptions, eghbal2020working}. Achieving this ideal was considered difficult when maintainers had reached different levels of success. For some, the feeling of joy and passion for their projects was more important. Although there was general agreement that the former could enable the latter, many expressed issues regarding work-life balance or even keeping an interest in their projects at all (cf.~\cite{eghbal2020working}). 

Being employed to work on the OSS project was considered a top scenario, e.g., to offer commercial support related to the project, to address the employer's internal needs, or as a means of driving personal development and education regularly. Creating and growing a business was the second option beyond employment, where some of the maintainers also had personal experience. Compared to employment, this option was associated with a higher personal risk for the maintainer but was also trickier in how to prioritize the needs of the community and the customers. A third option highlighted was sponsorship from individuals and companies. Sponsorship is percieved as less about sustainable availability of resources (referred to as ''pocket money'' or "coffee money' by our interviewees) and more as means of appreciation (aligning with~\cite{overney2020how, zhou2022studying}), employment and turning an open source project into a business were perceived as the more sustainable options, however, also more difficult to attain.

\begin{tcolorbox}\textbf{P5}: Maintainers who are financially supported to dedicate time to maintaining their OSS projects experience a higher level of work-life balance.
\end{tcolorbox}

\section{Limitations and Threats to validity}

Per our qualitative research approach, we use the four criteria for naturalistic inquiries proposed by Guba~\cite{guba1981criteria}.

\textbf{Credibility} regards the truth value of the presented findings. To strengthen this aspect, we performed member checking to enable each maintainer to correct, add or retract any statements in the transcripts. We also performed continuous discussions and peer debriefings to maintain situational awareness of our observations, and the perceived level of saturation.

\textbf{Transferability} concerns if and how the presented findings can be generalized to other cases beyond those studied. In this study, we focus how the availability of maintenance labor impacts the health of OSS projects from a maintainer perpective, and how the construct of human infrastructure can help to sustain this availability. We purposefully selected OSS projects with a limited number of maintainers to explore the perceptions about health and sustainability for projects of projects with limited corporate support. We do not attempt to generalize across any specific type of OSS projects, further recognizing that this is an exploratory and qualitative study. Readers, therefore, need to consider the context from where the data is collected to enable any anecdotal generalizations. Continuous use of quotes from the interviews has been used consciously to provide rich and thick contextualization of the findings, and each challenge or action specifically. 

\textbf{Dependability} refers to how reliable the results are in their replicability and trackability throughout the chain of evidence. To strengthen this aspect, we have maintained an audit trail tracing back throughout the data analysis and collection process. Also, we provide traceability for the questionnaire back to the original assumption through the assurance case, enabling a deeper understanding of the rationale for the research design, and thereby improving the replicability of the study.

\textbf{Confirmability} concerns to what extent presented findings were objectively derived. In this regard, we applied investigator triangulation~\cite{guba1981criteria} throughout the research design and implementation process to ensure neutrality and reduce the risk of introducing researcher bias. This included several steps to maintain the research focus and chain of evidence, from our initial assumption to the design of the questionnaire and the iterative coding process. The first two authors collaboratively designed the assurance case, and the derived questionnaire, both of which were externally reviewed and piloted. The coding process was performed individually by the first two authors, and discussions were facilitated by the third author who acted as a referee where deviating codings occurred.

\section{Conclusions and Future work}
Our study provides an exploratory investigation into how human infrastructure may help to support and secure the necessary amount of maintenance labor needed to ensure the long-term health of an OSS project. We identify a framework of infrastructure design aspects that maintainers and communities may consider when building human infrastructure. The design aspects consider specifically the human factors required, e.g., enabling a sustainable work-life balance, managing social pressure, promoting inclusiveness, and setting up guard rails for managing toxicity. The design aspects further considers the two categories of maintenance labor as \textit{Maintainer} and \textit{Contributor Labor}, representing labor coming from the maintainers and the contributors of an OSS project respectively. Both sources of labor are depletable and need to be replenished and sustained for an OSS project to stay healthy, i.e., viably maintained long-term without interruption or weakening.

Our observations suggest that there is a dependency between the two categories where a limited amount of Contributor Labor creates a burden and dependency on the maintainer, while too high amounts may create too large of an overhead for maintainers, resulting in increased stress and pressure. Attraction of new contributors is considered pivotal for a healthy project, but needs to be balanced with an onboarding process that can empower the newcomers to make better contributions, requiring less attention from the maintainers. Non-code contributions, such as helping out with support, community management, and marketing are specifically highlighted as important contributions. Finally, \textit{Resource Funding} is seen as a means of enabling maintainers and contributors to create a healthy work-life-balance, while also improving the project health.

Our findings are indicative by nature and need further exploration and validation in future work. Our findings provide guidance for maintainers, contributors, and consumers of OSS on how to reflect on and communicate how they can individually and collaboratively help to increase the health and sustainability of their respective OSS projects.


\bibliographystyle{ACM-Reference-Format}
\bibliography{bibliography_master}

\end{document}